\documentclass[useAMS,usenatbib]{mn2e}

\usepackage{aas_macros,amsmath,graphicx,lineno,longtable,amssymb,lastpage,mydefs}
\usepackage[version=3]{mhchem}

\title{A Minimum Mass Nebula for M Dwarfs}

\author[Gaidos]{E. Gaidos,$^{1,2}$\thanks{Fulbright Fellow}\\
$^{1}$Department of Geology \& Geophysics, University of Hawaii at M\={a}noa, Honolulu, Hawaii 96822 USA\\
$^{2}$Institute for Astrophysics, University of Vienna, Vienna, 1180 Austria\\
}

\begin{document}
\date{MNRAS, in press}
\pagerange{\pageref{firstpage}--\pageref{lastpage}} \pubyear{2017}

\maketitle

\label{firstpage}

\begin{abstract}
Recently revealed differences in planets around M dwarf vs. solar-type stars could arise from differences in their primordial disks, and surveys of T Tauri stars find a correlation between stellar mass and disk mass.  "Minimum" disks have been reconstructed for the Solar System and solar-type stars and here this exercise is performed for M dwarfs using \kep{}-detected planets.  Distribution of planet mass between current orbits produces a disk with total mass of $\approx$0.009\msun{} and a power-law profile with index $\alpha=2.2$.  Disk reconstruction from the output of a forward model of planet formation indicates that the effect of detection bias on disk profile is slight and that the observed scatter in planet masses and semi-major axes is consistent with a universal disk profile.  This nominal M dwarf disk is more centrally concentrated than those inferred around the solar-type stars observed by \kep{}, and the mass surface density beyond 0.02~AU is sufficient for in situ accretion of planets as single embryos. The mass of refractory solids within 0.5~AU is 5.6\mearth{} compared to 4\mearth{} for solar-type stars, in contrast with the trend with total disk mass.  The total solids beyond 0.5~AU is sufficient for the core of at least one giant planet.
\end{abstract}


\section{Introduction}

The exoplanet surveys conducted by ground-based radial velocity (RV) spectrographs and the NASA {\it Kepler} mission and its successor {\it K2} show that planets are ubiquitous around main sequence stars \citep{Howard2012,Burke2015}.  Moreover, analyses have found trends with stellar mass; the occurrence of Jovian planets probably increases with stellar mass \citep{Johnson2010,Gaidos2014}, while the occurrence of``small" (Earth- to Neptune-size) planets {\it decreases} \citep[e.g.,][]{Howard2012,Mulders2015a}.  The typical M dwarf star hosts 2 Earth-size or larger planets on orbits shorter than 0.5~yr \citep{Dressing2015,Gaidos2016}, a factor of $\sim$5 greater than solar-type stars \citep{Petigura2013}.  Understanding the origin of these differences is one test of planet formation theory and crucial for predicting other variation such as composition, atmospheres, and habitability in planet populations.

One important determinant of planet formation is likely to be initial conditions, i.e. the mass and structure of the protoplanetary disk.  Surveys of disks at sub-millimeter wavelengths show an overall positive relation between stellar and disk mass, either linear \citep{Andrews2013} or steeper \citep{Pascucci2016}.  However, this relation is significant only over a large mass range and the system-to-system scatter is comparable to the overall difference between G and M dwarfs predicted by the relation.  It may be obfuscated by variation in disk lifetime withs stellar mass \citep{Kastner2016} and the fact that the observations are only sensitive to dust and assumptions of grain size and lifetime.

A complementary approach is to reconstruct the mass and structure of protoplanetary disks using observed systems of planets.  This was first done for the Solar System \citep{Weidenschilling1977,Hayashi1985} and later extended to known exoplanet systems, initially the giant planets around solar-type stars detected by RV \citep{Kuschner2004}.  The masses of the planets are augmented to restore a solar-like composition and the mass spread over the intervening area between orbits.  This Minimum Mass (Exo)Solar Nebula is usually formulated as a power-law of mass surface density (gas or solids) with semi-major axis.  

\citet{Chiang2013} used the much larger and more complete catalog of \kep{} planets published by \citet{Batalha2013} to construct a MMEN.  They recovered a surface density up to $\approx$5 times that of the MMSN which would allow planets to accrete in situ.  However, there is dynamical evidence for planet migration in the Solar System \citep[e.g.,][]{Malhotra1993,Fernandez1996,Levison2008} and other systems \citep{Lin1996,Mills2016}, which would have redistributed mass.  \citet{Desch2007} included the migration of the giant planets described by the ``Nice" model to derive a steeper power-law profile for the MMSN.  \citet{Raymond2014} determined individual mass distributions for \kep{} multi-planet systems and found a wide range of power-law indices.  They interpreted this to show that the distributions are inconsistent with in situ accretion from a single, universal disk profile and that close-in planets migrated to their current orbits.

Previous reconstructions are primarily or exclusively for solar-type stars.  Here, reconstructions of the inner ($<$1~AU) disks around M dwarfs are performed using \kep{} statistics, extrapolated to $\gtrsim 1$~AU where RV and microlensing surveys probe, and compared to the results from solar-type stars.  

\section{Methods}
\label{sec.methods}

\begin{figure}
\centering
   \includegraphics[width=\columnwidth]{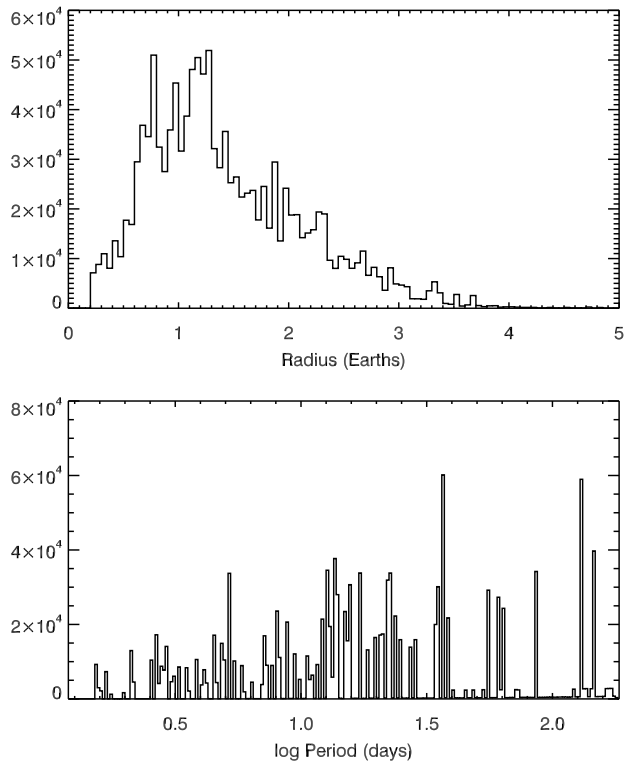}
   \caption{Distributions of planet radius (top) and orbital period (days) in the planet population inferred around \kep{} M dwarf stars \citep{Gaidos2016} and used to reconstruct the protoplanetary disks.  By construction, this synthetic intrinsic population is significantly larger than the observed population.  This method preserves and over-samples the values but not occurrence of the actual planet radii and periods.}
\label{fig.radper}
\end{figure}

{\it Planet population:} The starting point is the orbital period and planetary radius distributions inferred by \citet{Gaidos2016} from an analysis of the \kep{} observations of M dwarfs.  All red ($r-J > 2.2$) stars were classified as dwarf or giant stars based on available spectra and photometry.   The intrinsic, de-biased planet population with orbital period $P<180$~d was constructed using the method of iterative Monte Carlo \citep{Cappe2007}, where a synthetic input population of planets is evolved by repeated passage through a detection simulation \citep{Silburt2015}.  Host stars and planets are randomized with the assumption that, within the sample, their properties are uncorrelated.  Planets are treated as discrete objects and not binned, but the observed distributions of observed periods and radii are over-sampled to generate the intrinsic population, which can be arbitrarily large to thoroughly sample the error distributions.  Radius and orbital period distributions of this synthetic population are shown in Fig. \ref{fig.radper}; the scatter reflects the counting statistics of the much smaller {\it observed} population ($\sim 100$ planets) rather than the synthetic one.  

{\it Planet masses:} Masses $M_p$ were assigned to planets according to their radii, using the statistical relations of \citet{Chen2017} (See \citet{Wolfgang2016} for similar relations).  These relations are based on well-studied systems with masses established either by the radial velocity or transit timing methods; although a systematic difference in masses from the two methods from observational biases is expected \citep{Steffen2016}, this is small compared to the scatter and near zero for Earth-size planets \citep{Mills2017}.  Many planets, particularly those with radii $>1.6$\rearth{} have mean densities lower than that of an Earth-like planet with a metal core and silicate mantle \citep{Rogers2015}.  These objects must have thick, low mean molecular weight envelopes containing hydrogen and helium, but also H$_2$O, NH$_3$, and/or CH$_4$, that contribute significantly to the radius but less so to the mass.  The envelope mass must be subtracted before the composition is restored to the stellar composition.    

\begin{figure}
\centering
   \includegraphics[width=\columnwidth]{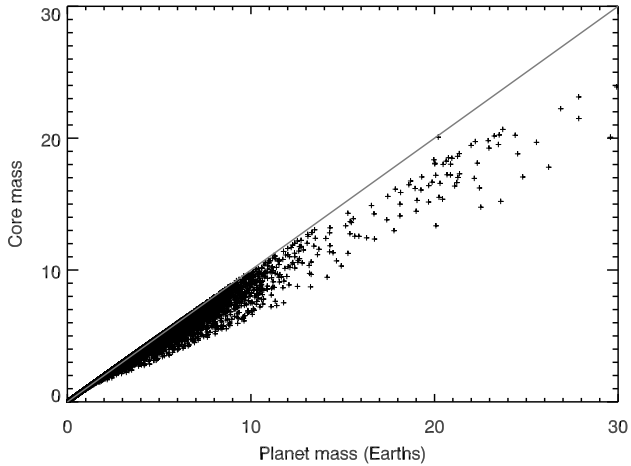}
   \caption{Predicted rocky core mass vs. total planet mass for the inferred planet population around \kep{} M dwarfs.  The smallest planets are almost exclusively rocky while most larger planets have appreciable envelopes dominated by H and He.}
\label{fig.mass}
\end{figure}

Each planet is assumed to consist of a rocky/icy core with an envelope of stellar composition and mean molecular weight $\bar{\mu}= 2.3$.  (The small correction for the non-solar stellar metallicity is neglected.)  The core mass is estimated by subtracting the mass of an accreted atmosphere $M_{core} = M_p - M_{atm}$, the latter calculated using the theory of \citet{Ginzburg2016}:
\begin{equation}
M_{atm} = \frac{\gamma-1}{\gamma}4\pi \rho_{rcb} R_c^2 R_{rcb} \left(\frac{R'_B R_{rcb}}{R_c^2}\right)^{1/(\gamma-1)},
\end{equation}
where $\gamma = 1.4$ is the adiabatic constant, $\rho_{rcb}$ and $R_{rcb}$ are the density and radius at the radiative-convective boundary, $R_c$ is the core radius, and $R'_B$ is the modified Bondi radius:
\begin{equation}
\label{eqn.bondi}
R'_B = \frac{\gamma-1}{\gamma} \frac{GM_C\bar{\mu}}{k_BT_{rcb}}.
\end{equation}
$\rho_{rcb} = p_{rcb}\bar{\mu}/(kT_{rcb})$, where $p_{rcb} \sim  0.1$~bar, a typical value for planetary atmospheres \citep{Robinson2012}, $R_{rcb} \approx R_p$, and $T_{rcb}$ was set to the planet's equilibrium temperature $T_{eq}$.  The mass-radius relation for the core was expressed as a power law $R_c = R_0 (M_c/M_0)^{\beta}$ with $\beta = 0.26$ \citep{Valencia2007}.  \teq{} was calculated assuming an albedo of 0.1.  Figure \ref{fig.mass} shows the predicted core mass vs. planet mass for the inferred planet population.  These masses were then restored to the stellar composition using the mass fraction in silicates and metals in the Solar System  \citep[0.0049,][]{Lodders2003}, adjusted for stellar metallicity.  

{\it Disk reconstructions:} The first reconstruction method computed the cumulative distribution of $M_p$ with semi-major axis, and then calculated the differential mass surface density $\Sigma$ using two methods; piecewise fitting of polynomials to the cumulative distribution followed by analytical derivation of the derivatives, and numerical calculation of the derivative of a running median ($N=99$).  For the first method, the fitting successively bifurcated the distribution, fitting each moiety with a quartic function and finding the bisector location that minimizes the square of the deviations of the fit.  

The second method assumed that each planet accreted as a single, isolated protoplanet and the local surface density was calculated by assuming the mass of the planet is spread over its feeding zone \citep{Schlichting2014}, i.e.
\begin{equation}
\label{eqn.embryo}
\Sigma = \frac{M_*^{1/3}M_p^{2/3}}{2^{5/2}\pi a^2}\left(\frac{R_*}{a}\right)^{-1/2}\left(\frac{\rho}{\rho_*}\right)^{-1/6},
\end{equation}
where $M_*$, $R_*$, and $\rho_*$ are the stellar mass, radius, and density, $a$ is the semi-major axis, $\rho$ is the planetesimal density ($\approx$3~g~cm$^{-3}$).  This procedure produced values at discrete locations and a running median ($N=99$) was calculated. 

The third method assumes inward migration of planets to their present orbits.  Planet mass was set to the mass in the disk between the planet's initial ($a'$) and final ($a$) orbit.  The surface density is assumed to follow a power-law profile $\Sigma \propto a^{-\alpha}$, and the mass of the $i$th planet is
\begin{equation}
m_i = \int_{a_i}^{a'_i} 2\pi \Sigma_1 a^{1-\alpha} da.
\end{equation}
Then the surface density is
\begin{equation}
\Sigma(a) = \frac{f}{N}\frac{\alpha-2}{2\pi} \sum\limits_{a_i < a < a'_i} \frac{m_i}{a_i^2}\left(\frac{a}{a_i}\right)^{-\alpha}.
\end{equation}
The initial orbit is assumed to be the current location of the {\it next} outer planet.  To calculate $a'_i$, orbits are assumed to be uniformly distributed with $\log a$ such that $a'_i = a_i (1-X)^c$, where $X$ is a random uniform deviate and $C$ is set so that there are 2.2 planets with $1.5<P<180$~d \citep{Gaidos2016}.  The summation is normalized by the number of systems and the number of planets per system.  $\alpha$ was determined by iteratively fitting a power law to the profile.

\section{Results}
\label{sec.results}
\begin{figure}
\centering
   \includegraphics[width=\columnwidth]{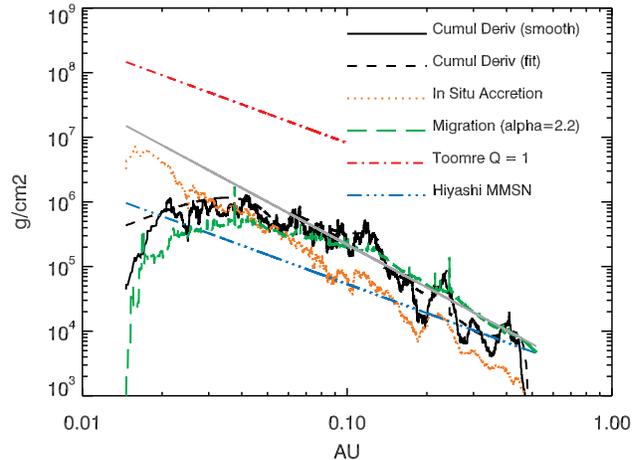}
   \caption{Reconstructed gas + solids mass surface density profiles for M dwarf disks.  The black solid and dashed lines are based on the cumulative distribution of planet mass vs. semi-major axis.  The dotted orange line is the requirement for in situ accretion of the planets, and the dashed green line assumes accumulation of disk solids as each planet migrates to its current orbit.  The dash-dot red line is the Toomre $Q=1$ condition for disk instability, and the dash-multi-dot blue line is an inward extrapolation o f the \citet{Hayashi1985} MMSN.  The grey line is a best power-law fit to the migration profile.}
\label{fig.disk}
\end{figure}

The reconstructed disk profiles in different scenarios are plotted in Fig. \ref{fig.disk}.  The reconstructions fail interior to 0.02~AU and exterior to 0.5~AU due to the paucity of planets at those locations (the input planet population is limited to $P<180$~d).  A best-fit power-law of the surface density profile beyond 0.06~AU derived under the migration scenario returns $\alpha = 2.2$ and $\Sigma_1 = 1380$~g~cm$^{-2}$ (gas plus solids) at 1 AU.  This is plotted as the grey line in Fig. \ref{fig.disk}.  Extrapolated without limit, the total mass in the disk is $\approx9\times 10^{-3}$\msun{}.  Because $\alpha$ is close to 2, this figure does depend if the disk has an outer edge.  This disk is ``super-Hayashi" inside of 0.75~AU (where the \kep{} planets orbit) but its extrapolation to greater distances is ``sub-Hayashi".

\begin{figure}
\centering
   \includegraphics[width=\columnwidth]{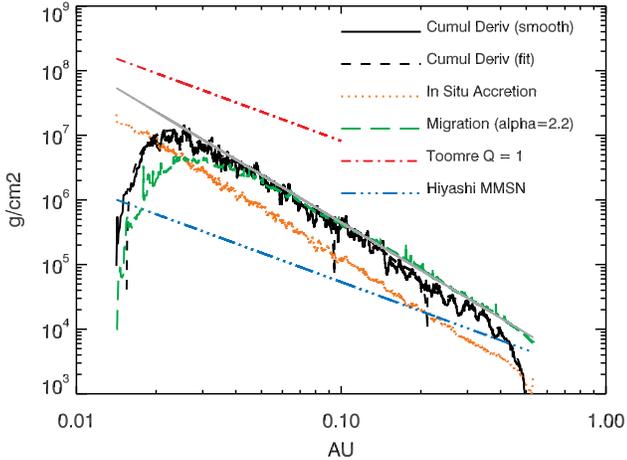}
   \caption{Reconstructions of disk mass surface density using a {\it simulated} planet population produced by modeling migrating accretion from a universal power-law disk with index $\alpha = 2.2$.  Compare to Fig. \ref{fig.disk}.}
\label{fig.sim}
\end{figure}

\kep{} detected only a small fraction of all planets around M dwarfs, and to address the possible effects of detection bias on disk reconstruction, a forward simulation was performed.   This selected M dwarf host stars from the catalog of \citet{Gaidos2016} and assigned them disks with power-law profiles with specified index $\alpha=2.2$.  The fraction of disk solids was scaled with the metallicity of the star.  Orbital semi-major axes were selected at 0.02~AU with each successive orbit using the distribution described in Sec. \ref{sec.methods} to achieve a uniform distribution with $\log a$ normalized to the mean number of planets with $P<180$~d.  Planet masses were set to the mass in the disk between orbits (i.e. migration assumption), radii were calculated using the \citet{Chen2017} relations, and transit depths were calculated using the host star radius.  Mutual orbital inclinations were selected from a Rayleigh distribution with $\sigma = 0.2$~deg \citep{Gaidos2016} with uniformly-distributed nodes of ascension.  The transit duration was calculated assuming zero impact parameter and the photometric noise over the duration was interpolated from the established values for the \kep{} Combined Differential Photometric Precision.  The total transit signal-to-noise ratio (SNR) was calculated as the ratio of the individual transit depth divide by the noise, times the square root of the expected number of transits during the quarters the star was observed.  The criteria for inclusion in the synthetic catalog were a transiting orbit, at least three transits in all observed quarters, SNR $>7.2$, and $P<180$~days.  Disk profiles were then reconstructed as in Sec. \ref{sec.methods}.  The derived profiles (Fig. \ref{fig.sim}) closely resemble those from the actual planet population (Fig. \ref{fig.disk}).  A power-law fit to the migration scenario profile returned $\alpha = 2.4$, suggesting a slight steepening like that observed by \citet{Raymond2014}.  A second simulation with $\alpha = 2$ yielded $\alpha = 2.2$, suggesting that the former value is closer to the true one.

\section{Discussion}
\label{sec.discussion}

Reconstruction of disks around M dwarfs based on the de-biased planet population detected by \kep{} indicate the following: (1) The restored mass surface density $\Sigma$ derived either by ``smoothing" planet masses over the present orbits or by assuming the planets represent the mass in the disk swept up by migration to their current orbits are similar; (2) The best power-law fit for the total surface density beyond 0.06~AU is $\Sigma = 1380\, a_{\rm AU}^{-2.2}$~g~cm$^{-2}$ and, for solar abundances, a refractory surface density $\Sigma_{\rm ref} = 6.7\,a_{\rm AU}^{-2.2}$~g~cm$^{-2}$, although the de-biased index may be closer to $\approx$2; and (3) The total extrapolated disk mass is $\approx 0.009$\msun, or 53\mearth{} in total (silicates + metals + ices) condensates.  The \kep{} survey was magnitude limited and as a consequence the median estimated mass of this sample is  0.46\msun{}, near the high end for M dwarfs.  The recently discovered TRAPPIST-1 system \citep{Gillon2017} suggests that radially concentrated mass distributions may extend to the less luminous but more numerous lower-mass M dwarfs. 

This M dwarf disk profile is more centrally concentrated than the canonical MMSN of \citet{Hayashi1985} ($\alpha = 1.5$) and that derived by \citet{Chiang2013} for solar-type \kep{} stars ($\alpha = 1.6$) but is similar to the migration-corrected MMSN of \citet{Desch2007}, as well as that of RV-detected exoplanet systems \citep{Kuschner2004}.  The disk mass is at the upper end of the range of observed M dwarf disk masses \citep{Andrews2013,Pascucci2016} but below the canonical MMSN value of 0.013\msun.  The average mass of refractory solids inside 0.5~AU is 5.6\mearth.  The disk derived by \citet{Chiang2013}, adjusted by a ratio of refractory to total condensates of 0.32, contains 4.1\mearth.  These values are slightly lower than the values of 5 and 7\mearth{} found by \citet{Mulders2015b} for $P<50$d.  

Some M dwarfs host Neptune- to Jupiter- size planets on more distant orbits that were poorly sampled by \kep{} but do appear in RV and microlensing surveys.  \citet{Gaidos2014}, using both RV and \kep{} statistics, found that  $\approx$3\% of solar-metallicity M dwarfs have planets more massive than $0.3 M_J$ to $P = 2$~yr.  Metal-rich M dwarfs are also more likely to host giant planets \citep{Mann2013,Neves2013,Gaidos2014}.  \citet{Clanton2014}, using RV and microlensing surveys, estimated the {\it total} occurrence of $>$0.16\mjupiter{} planets to be $\approx$11\%.  Since the disks inferred here are stable against self-gravity (Toomre $Q \ll 1$, Fig. \ref{fig.disk}), giant planets would have to form by core-first accretion, requiring a solid core $\approx$10\mearth{} \citep{Piso2015}.  These disks contain $\approx$7~\mearth{} of refractory solids inside a 2-yr orbit.  The mass further out, including H$_2$O ice in Solar System abundance \citep{Lodders2003}, is $\approx$16\mearth.  Thus there is adequate mass to form a core, and sufficient gas ($\sim 9M_J$) to form a giant planet envelope.  Giant planet-hosting M dwarfs are more likely to be metal rich and the inventory of refractory solids would be double ([Fe/H $>0.3$) in 4\% of \kep{} M dwarfs, according to the metallicities estimated by \citet{Gaidos2016}.  These inventories do not address the issue of {\it accretion time}, an acute problem if M dwarfs disks dissipate more rapidly \citep{Kastner2016}.

The nominal M dwarf disk derived here has a sufficient $\Sigma_{\rm ref}$ at $a > 0.05$ for planets to accrete in situ from single embryos, as proposed by \citet{Chiang2013} for \kep{} exoplanets orbiting solar-type stars.  Planets inside 0.05~AU would have formed further out and migrated inwards, since growth via collisions of multiple embryos would be inhibited by the low Safronov number ($<0.01$).  The solids need not be mirrored by an equal concentration of gas, and indeed this is unlikely since efficient migration in such a disk would efficiently remove planets \citep{Crida2009}.  Instead, solids may have drifted inwards and concentrated due to gas drag.    

\begin{figure}
\centering
   \includegraphics[width=\columnwidth]{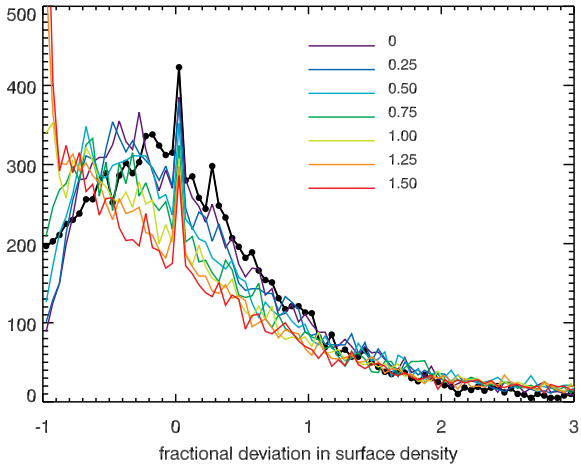}
   \caption{Distributions of the fractional deviation in surface density (estimated assuming in situ planetary accretion) from a running median value with semi-major axis.  The black line with points is the observed distribution.  The colored lines are for simulated planets formed from power-law disks with mean index $\alpha = 2.2$ and different standard deviations.}
\label{fig.hist}
\end{figure}

\citet{Raymond2014} argued against a {\it universal} minimum-mass extrasolar nebula (for solar-type stars) based on the diversity of profiles derived from multi-planet \kep{} systems.  They concluded that planet migration produced this variation and therefore that planets did not accrete in situ.  Randomization by the Monte Carlo method used to infer the intrinsic M dwarf planet population precludes this system-by-system approach,  but the {\it scatter} in estimates of disk density, i.e. assuming in situ accretion, can be determined and compared to simulated systems arising from a universal disk profile.  Surface densities were calculated using the forward model described in Sec. \ref{sec.results} and disks with different profiles described by a normal distribution of $\alpha$ with mean of 2.2 and a specified standard deviation $\sigma_{\alpha}$.  A running median ($N=99$) curve was subtracted and the fractional dispersion calculated.  Distributions of the deviations for the actual population and simulations with different $\sigma_{\alpha}$ are plotted in Fig. \ref{fig.hist}.  The $\sigma_{\alpha} = 0$ case (purple) most closely follows that of the actual distribution black.  The Tukey's biweight robust standard deviations are 0.77 and 0.72, respectively. Simulations with increasing $\sigma_{\alpha} > 0$ have distributions that increasingly deviate from the observations. This shows that random processes in this model other than variation in disk profile (e.g., orbital periods) are sufficient to explain the observed scatter, but of course, the model may not be unique.

\end{document}